\documentclass{article}
\usepackage{multicol}
\usepackage[utf8]{inputenc}
\usepackage[margin=1in]{geometry}
\usepackage{booktabs}
\usepackage{multirow}
\usepackage{graphicx}
\usepackage{lscape}
\usepackage{placeins}
\usepackage{authblk}
\usepackage[table,xcdraw]{xcolor}

\title{Emerging Trends of Recently Published Datasets for Intrusion Detection Systems (IDS): A Survey}

\author{Rishabh Jindal $^{1}$ and Adnan Anwar $^{2}$}
\affil{%
$^{1}$ \quad School of IT, Deakin University\\
$^{2}$ \quad Centre for Cyber Security Research and Innovation, School of IT, Deakin University}

\date{\vspace{-5ex}}

\begin{document}

\maketitle

\rule{\columnwidth}{0.5pt}
\begin{abstract}
\noindent With the ubiquitous nature of information technology solutions that facilitate communication in the modern world, cyber attacks are increasing in volume and becoming more sophisticated in nature. From classic network-based Denial of Service (DoS) attacks to the more recent concerns of privacy compromises, Intrusion Detection Systems (IDS) are becoming an urgent need to safeguard the modern information technology landscape. The development of these IDS relies on training and evaluation datasets that must evolve with time and represent the contemporary threat landscape. The purpose of this analysis is to explore such recent datasets, describe how they enable research endeavours and the development of novel IDS. Specifically, 7 recent datasets published for IDS research have been reviewed along with selected publications that have employed them. In doing so, the discussion emphasizes the need for the publication of even more modern datasets, especially for emerging technologies such as the Internet of Things (IoT) and smartphone devices, to ensure that modern networks and communication channels are secured. Furthermore, a taxonomy based on the discussed datasets has been developed that can be used to inform the dataset selection process for future research endeavours.
\end{abstract}
\rule{\columnwidth}{0.5pt}

\section{Introduction}
Given the increasing digital footprint of human existence; even more so in the aftermath of the Covid-19 pandemic, the threat posed by cyber attacks is greater than ever. The cyber-attack landscape is constantly evolving; with the number and types of attacks increasing regularly. In a recent survey, researchers identified over 50 ill consequences of cyber attacks across 5 broad categories – Physical/Digital, Economic, Psychological, Reputational and Social/Societal~\cite{agrafiotis2018taxonomy}. 

Such attacks can not only lead to individual and economic losses but can cause a great deal of damage to society at large. An example of this is the attack on Ukraine’s power grid system in 2015. Attackers were able to gain access to the system, control over 30 substations and leave over 230,000 people without light or heat in the peak of winter~\cite{zetter_2016}. Therefore, it is imperative to develop tools that can effectively identify even sophisticated attacks and notify the appropriate experts in a timely fashion.

Intrusion Detection Systems (IDS) are one set of such tools. The primary goal of IDS is to monitor networks and identify malicious intrusions. These systems can broadly be classified into two categories based on how they identify intrusions – signature-based methods scan the network for characteristics of known attacks, whereas anomaly-based methods scan the network for traffic that deviates from the usual nature of traffic on that network. There are also hybrid approaches that can combine strengths of both signature and anomaly-based methods. 

The modern information technology landscape is constantly evolving and so is the nature of communication being facilitated by devices across the globe.  With the advent of smart devices and the internet of things, we live in a world where the number of networked devices far exceeds the number of people in the world. However, novel research solutions in the area of intrusion detection are still disproportionately evaluated on traditional datasets such as NSL-KDD, DARPA, etc, that do not accurately represent the nature of threats in the modern IT landscape.  This is in large part due to the lack of modern datasets that contain a representative sample of attacks, especially for newer technologies such as the internet of things.

In this survey, we present a discussion of seven recently published modern datasets and the research endeavours that they have enabled in an effort to highlight the need for more such datasets to ensure that the intrusion detection research community can continue to thrive and be able to nip even the most sophisticated threats in the bud.

Specifically, the highlights of the current work are that it provides a detailed discussion of recent datasets and their importance in the advancement of IDS research, and the development of a taxonomy for IDS datasets. The rest of this analysis is structured as follows: section 2 presents a discussion of related surveys; section 3 presents a brief overview of IDS; section 4 first describes the dataset selection procedure and then presents the discussion of the selected datasets and the research informed by them; section 5 presents a taxonomy of the selected datasets and finally, concluding remarks as well as directions for future research are presented in section 6.

\section{Related surveys}
This paper adopts a systematic approach to the review of intrusion detection datasets and includes a wide range of datasets that have served as the benchmark in research over recent years. Although most intrusion detection research has leveraged older and more traditional datasets, this paper focuses on discussing recent datasets – published in or after the year 2017, to emphasize the importance of creating new datasets that reflect the modern threat landscape and the research opportunities that are created by the publication of such modern datasets. Most related surveys that address intrusion detection systems focus primarily on the algorithms used and/or developing taxonomies of these techniques and the datasets used. The present study is different in its focus on the study of recent datasets and their suitability for use in the modern IT environment. The specific contributions of this paper are described below: 

\begin{enumerate}
\item One of the major aims of this survey is to examine the emerging datasets that have been published in recent years (in or after the year 2017). A comprehensive discussion regarding the development and suitability of each dataset is undertaken to emphasize the value delivered by modern datasets and the need for more such modern and representative datasets to drive future research endeavours.
\item The study also aims to develop a comprehensive taxonomy of these IDS datasets. While existing taxonomies have focussed on internet and network traffic data, this study broadens the scope and includes datasets from the fields of IoT and smartphone malware to highlight the difference in the underlying characteristics between categories and the need for custom-tailored IDS solutions for each.
\end{enumerate}

The 2016 survey by Buczak et al.~\cite{buczak2015survey} presented an effective overview of intrusion detection and the most popular machine learning methods used for the same. The survey’s focus is to educate the reader about the fundamentals of intrusion detection, machine learning and the metrics used to evaluate the performance of various algorithms. Regarding datasets, a brief discussion is presented that provides only a description of the datasets that are used in the algorithms discussed in the study. Therefore the present study is different in its focus on delivering a detailed discussion regarding the development of IDS datasets in recent years and the need for more such datasets to be published, rather than the various machine learning and deep learning algorithms that are built on top of them.

In the 2019 paper by Khraisat et al.~\cite{khraisat2019survey}, the authors present a taxonomy of different IDS techniques, focusing primarily on the underlying machine learning techniques with only a brief discussion of some popular network traffic datasets irrespective of their year of publication. In contrast, the current survey presents a taxonomy of the recent datasets that are used to study IDS techniques, broadening the scope to include datasets from the IoT and smartphone malware fields in addition to the widely explored field of network traffic. Furthermore, by focussing on recent datasets, this study illustrates the increased validity and wider applications of such datasets compared to older datasets that are becoming increasingly obsolete, and emphasizes the need for the publication of even more modern datasets.

The 2020 study by Thakkar et al.~\cite{thakkar2020review} is focussed on IDS datasets and studies the recent CIC-IDS-2017 and CSE-CIC-IDS-2018 datasets to analyse how IDS datasets have evolved over time. The present study builds on this effort and analyses more popular recent IDS datasets, highlighting the evolution of such datasets and their increased validity and reliability to evaluate the performance of modern mitigation techniques in real-world scenarios.

The 2019 study by Ring et al.~\cite{ring2019survey} provides one of the most comprehensive reviews of IDS datasets that the authors came across during their literature review. The study analyses 34 popular network traffic-based datasets based on 15 parameters. The authors explore the structure and generation of each dataset to identify the best-suited scenarios for its use and its limitations. The present study builds on this analysis by analysing the situations in which such modern datasets have actually been deployed and how the publication of such datasets can inform research endeavours in unexpected ways and thus lead to the rapid advancement of the field.

The 2020 study by Hindy et al.~\cite{hindy2020taxonomy} presents taxonomies for IDS techniques, datasets and threats. Since the study concerns itself with building taxonomies, the analysis of datasets is primarily a statistical one that uncovers the distribution and frequency of various datasets in the literature and the nature of attacks represented in those datasets. Therefore, the present study is different in its greater focus on the study of IDS datasets and its evaluation of the suitability and reliability of these datasets in the modern intrusion threats environment.

The 2020 survey conducted by Aldweesh et al.~\cite{aldweesh2020deep} was more focused and provided a taxonomy of deep learning-based IDS techniques with respect to factors such as “input data, detection, deployment and evaluation strategies.” One of the conclusions that the authors reached was that the currently prevalent datasets are either obsolete because they do not capture the wide range of modern attacks or do not effectively simulate real-world environments. The current survey aims to shed more light on this issue from another angle by discussing the suitability and reliability of recently-developed datasets from a real-world perspective.

The 2020 study by Ferrag et al.~\cite{ferrag2020deep} is another comprehensive study on IDS datasets. It describes 35 popular IDS datasets and creates a taxonomy based on them, proposing 7 categories under which IDS datasets can be classified. The study then analyses the performance of 7 deep learning techniques on 2 recent datasets (CSE-CIC-IDS2018 and BoT-IoT) to evaluate the efficacy of deep learning techniques for IDS. The present study builds on the study of IDS datasets initiated by Ferrag et. al by evaluating the validity and reliability of recent datasets to evaluate the performance of modern IDS techniques. Specifically, the present analysis conducts a thorough investigation of recently developed IDS datasets to evaluate whether they are representative of the wide gamut of modern intrusion attacks that are encountered in the real world.

\begin{table}[]
\centering
\resizebox{\textwidth}{!}{%
\begin{tabular}{@{}|l|c|c|c|c|@{}}
\toprule
                        & \multicolumn{4}{c|}{Features}                                                                                                \\ \cmidrule(l){2-5} 
\multirow{-2}{*}{Paper} & Focus on recent datasets    & Ideal use-case of each dataset & Validity of   datasets          & Taxonomy of   datasets      \\ \midrule
Buczak et al.           & \cellcolor[HTML]{FBE4D5}No  & \cellcolor[HTML]{FBE4D5}No     & \cellcolor[HTML]{FBE4D5}No      & \cellcolor[HTML]{FBE4D5}No  \\ \midrule
Khraisat et al.         & \cellcolor[HTML]{FBE4D5}No  & \cellcolor[HTML]{FBE4D5}No     & \cellcolor[HTML]{FBE4D5}No      & \cellcolor[HTML]{FBE4D5}No  \\ \midrule
Thakkar et al.          & \cellcolor[HTML]{E2EFD9}Yes & \cellcolor[HTML]{FBE4D5}No     & \cellcolor[HTML]{FFF2CC}Partial & \cellcolor[HTML]{FBE4D5}No  \\ \midrule
Ring et al.             & \cellcolor[HTML]{FBE4D5}No  & \cellcolor[HTML]{E2EFD9}Yes    & \cellcolor[HTML]{FFF2CC}Partial & \cellcolor[HTML]{E2EFD9}Yes \\ \midrule
Hindy et al.            & \cellcolor[HTML]{FBE4D5}No  & \cellcolor[HTML]{FBE4D5}No     & \cellcolor[HTML]{FBE4D5}No      & \cellcolor[HTML]{E2EFD9}Yes \\ \midrule
Aldweesh et al.         & \cellcolor[HTML]{FBE4D5}No  & \cellcolor[HTML]{FBE4D5}No     & \cellcolor[HTML]{FBE4D5}No      & \cellcolor[HTML]{FBE4D5}No  \\ \midrule
Ferrag et al.           & \cellcolor[HTML]{E2EFD9}Yes & \cellcolor[HTML]{FBE4D5}No     & \cellcolor[HTML]{E2EFD9}Yes     & \cellcolor[HTML]{E2EFD9}Yes \\ \midrule
\textbf{Present study} &
  \cellcolor[HTML]{E2EFD9}\textbf{Yes} &
  \cellcolor[HTML]{E2EFD9}\textbf{Yes} &
  \cellcolor[HTML]{E2EFD9}\textbf{Yes} &
  \cellcolor[HTML]{E2EFD9}\textbf{Yes} \\ \bottomrule
\end{tabular}%
}
\caption{Summary of related surveys and present study.}
\label{tab:caption}
\end{table}

\section{Overview of Intrusion Detection}
Intrusion detection refers to the process of monitoring a network or system to identify malicious activity. The set of tools and techniques used for this purpose are known as Intrusion Detection Systems (IDS). There are multiple types of IDS, as will be discussed shortly, however, the principle underlying all of them can be described as follows: an IDS collects considerable amounts of data by logging all the traffic on a system or network and attempts to identify malicious activity by cross-referencing such data across multiple sources. Intrusion Detection Systems can be classified into three main categories based on the technique used to identify malicious traffic.

\subsection{Signature-based detection}
A signature refers to the characteristics of a previously identified attack. Therefore, this technique aims to identify malicious activity by comparing traffic data to the signatures of previously identified attacks. However, this technique, also known as misuse detection, presents several challenges:

Since it relies on signatures of previously identified attacks, it cannot be to used identify novel or zero-day attacks. It is often cumbersome to program the signatures of all previously identified attacks, rendering the deployment of signature-based methods time-consuming and expensive.

\subsection{Anomaly-based detection}
An anomaly refers to a departure from normal behaviour, therefore, anomaly-based intrusion detection techniques aim to learn the traffic patterns of a given system or network and identify malicious activity by analysing any deviations from that normal pattern. The three primary strategies used to identify anomalies are described below:

\begin{itemize}
\item Statistics-based techniques: Such techniques strive to identify anomalous activity by analysing statistical indicators such as standard deviations, means and distributions.
\item Knowledge-based techniques: Such techniques strive to identify anomalous activity based on previous knowledge regarding normal and abnormal activity.
\item Machine and Deep learning techniques: Such techniques involve the use of mathematical algorithms to learn normal traffic patterns from past data without human intervention.
\end{itemize}

It is also important to note that since anomaly-based techniques do not rely on previously encountered attacks, they can be used to identify novel or zero-day attacks.

\subsection{Hybrid detection}
Signature-based methods suffer from shortcomings such as their inability to identify new attacks and a lack of dynamism owing to their hard-coded nature. On the other hand, anomaly-based methods pose their own set of problems such as the high rate of false positives. Hybrid techniques aim to effectively combine signature-based and anomaly-based detection techniques such that each technique can be used to mitigate the shortcomings of the other. Therefore, these techniques can offer better performance against novel attacks when compared to signature-based methods and produce a lower rate of false positives when compared to anomaly-based techniques.

\subsection{Deep learning vs. traditional machine learning for intrusion detection}
Arthur Samuel, often credited as the father of machine learning, defined it as a “field of study that gives computers the ability to learn without being explicitly programmed”~\cite{samuel1959some}. Machine learning primarily focusses on regression and classification tasks, thus making it an appropriate tool for the purpose of intrusion detection because it is in essence a classification problem. In traditional machine learning methods, the data that is used to train various models must be curated by experts and in the case of classification tasks, distinctly labelled observations must be provided. This can become a time-consuming and arduous procedure when large datasets like the ones observed in the modern information technology environment are encountered. This is where modern deep learning techniques enter the frame and offer considerable benefits.

One of the foremost texts in the field describes deep learning as “a particular kind of machine learning that achieves great power and flexibility by learning to represent the world as nested hierarchy of concepts, with each concept defined in relation to simpler concepts, and more abstract representations computed in terms of less abstract ones”~\cite{Goodfellow-et-al-2016}. In other words, deep learning architectures have the ability to learn the relationships between various features of the dataset through the use of abstractions and multiple layers and without the need for human intervention. Therefore, such architectures are able to address one of the biggest challenges of traditional machine learning methods, that is the need to provide models labelled training data.

Furthermore, from an intrusion detection perspective, deep learning architectures are well suited to the dynamic and large-scale nature of datasets in the modern information technology environment and can identify complex relationships between variables that would be impossible for humans to discern due to sheer scale and dimensionality. Therefore, deep learning models are ideal tools to deploy anomaly-based intrusion detection techniques through. They can analyse large amounts of historical data and classify traffic as either normal or anomalous and even further classify anomalous traffic into various categories of malicious attacks.

Although the theory behind deep learning has been around for decades, such techniques have historically been largely impractical due to the sizable requirement of computational resources. However, recent advances in hardware technology have made deep learning more accessible than ever before and thus there has been an explosion of interest in the field in recent years with innovative and remarkable applications across a myriad of domains.

\section{Recent datasets developed for intrusion detection research}

\subsection{Systematic Approach for Dataset selection}
As stated earlier in our discussion, the present analysis is concerned with the exploration and evaluation of recent datasets published for intrusion detection research. During the literature review process, the authors noted that there was a significant uptick in the number of such datasets being published in the year 2017 and after. Therefore, 2017 was selected as the cutoff year to limit the scope of the analysis to a manageable duration. The exact process that lead to the shortlisting of the 7 datasets that form the basis of the discussion is described below.

In addition to the widely researched area of network traffic intrusion detection, the researchers wanted to include discussions about areas that are gaining traction and represent an increasing amount of traffic in the modern information technology landscape: IoT devices and smartphones. Furthermore, the authors also wanted to include a discussion regarding the research in the area of privacy intrusion. As more and more people begin using technology in their daily lives and generate the vast amounts of data that we observe today, the threat to users' privacy is perhaps greater than it has ever been before. Therefore, privacy preservation is an active area of research in the intrusion detection community and the publication of more comprehensive datasets is an indispensable requirement for the advancement of the field. 

Finally, the authors turned to some of the most reputable institutions in the intrusion detection research community; such as the Canadian Institute for Cybersecurity, the Center for Applied Internet Data Analysis and UNSW Canberra, to explore which of their recent publications meet the outlined criteria. These institutions have a history of developing reliable and influential datasets in the past that have informed research endeavours for many years and have had a significant contribution in the evolution of the industry. Trusting the pedigree of these institutions was highly successful as the analysed datasets are comprehensive, reliable and have already formed the basis of many exciting and innovative research endeavours. This is highlighted by our detailed discussion of each selected dataset in the next section.

\subsection{Recent Datasets}

\subsubsection{TOR-nonTOR}
\textbf{Overview}: With the rapid rate of advancement in information technologies and the widespread adoption of the Internet across the world, the risk of privacy breaches is greater than ever before. Privacy is an active research area and has attracted the interest of many cybersecurity researchers. As with intrusion detection research, there is a pressing need for reliable and valid datasets that can be used to train and evaluate privacy preserving and compromising techniques. One such dataset was published in 2017 by the researchers in~\cite{lashkari2017characterization}. The dataset contains labelled traffic over the Tor network and was developed as part of the researchers’ effort to develop a technique to identify Tor traffic and classify it into one of eight categories – browsing, chat, audio-streaming, video-streaming, mail, VOIP, P2P, and File Transfer. Tor is currently one of the most popular tools used by internet users to encrypt and anonymize their traffic to avoid surveillance on the internet. Identifying Tor traffic and being able to classify it into one of several categories as suggested by the researchers represents a severe privacy risk for unsuspecting users. Thus, the researchers in~\cite{lashkari2017characterization} hope to increase the level of privacy that can be achieved on the Tor network by publishing this dataset to facilitate the study of its current vulnerabilities. The dataset generation process is discussed next.

\begin{figure}[htp]
    \centering
    \includegraphics[width=\columnwidth]{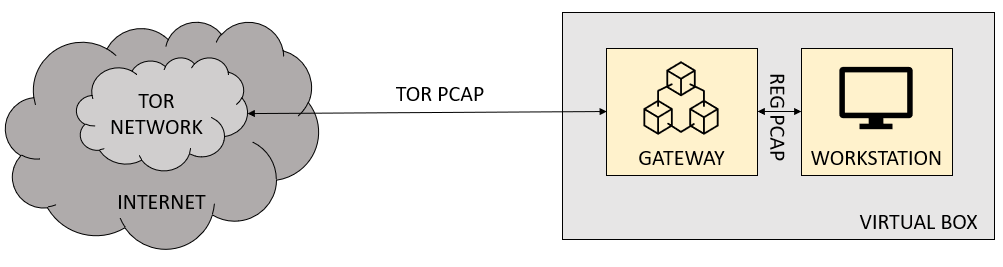}
    \caption{Architecture of data collection system}
    \label{fig:tor_nontor}
\end{figure}

The researchers set up a virtual box comprised of a workstation and a gateway to generate traffic across 8 categories (browsing, chat, audio-streaming, video-streaming, mail, VOIP, P2P, and File Transfer) over 18 relevant applications (Facebook, Spotify, Gmail, Skype, etc.). The traffic flows between the workstation and the gateway were represented by regular pcap files and the traffic flows between the gateway and the internet were represented by Tor pcap files. The Tor pcap files were then labelled based on sending only one application’s traffic (regular pcap files) across the workstation and gateway at a time.The system architecture is illustrated in Figure~\ref{fig:tor_nontor}.

\textbf{Related Literature:}
Even though Tor is primarily used by users to avoid being tracked and being served targeted advertisements, it has also been employed by criminals as a means of conducting illegal activities online without being traced. Therefore, techniques that can identify users on the Tor network may decrease the level of privacy for an average user, yet they are an essential tool in the arsenal of cybersecurity and law enforcement professionals to protect the interests of society at large. \cite{gurunarayanan2021improving} uses the Tor-nonTor dataset to evaluate and optimize the performance of several machine learning techniques with respect to Tor traffic identification. Specifically, the researchers employed random oversampling and under sampling to mitigate the class imbalance of the dataset and techniques such as K-fold cross-validation and Grid Search to optimize the hyperparameters of the selected machine learning algorithms. Based on their experiments, the researchers obtained the best results with the Random Forest algorithm and were able to achieve an accuracy of 98.68\% in identifying Tor traffic.

Another application of the Tor-nonTor dataset, that has been studied by researchers in~\cite{hodo2017machine}, is identifying nonTor traffic on a Tor network to shed light on the activities that when carried out on a Tor network can downgrade privacy safeguards and thus put users at risk.  Specifically, the study evaluated the performance of Artificial Neural Networks (ANN) and Support Vector Machines (SVM) at identifying nonTor traffic. Based on their experiments, the researchers concluded that an ANN built on features derived through correlation-based feature selection achieved the best results with a detection accuracy of 99.8\% for nonTor traffic. The discussed applications emphasize the versatility of the TOR-nonTOR dataset through the fact that it can be used to inform research endeavours both to increase as well as decrease the privacy-level of users on the Tor Network. As highlighted by our discussion thus far, both applications can be important based on the context being studied.

\subsubsection{CICAndMal2017}
\textbf{Overview:} This dataset was published in 2017 in response to the lack of representative datasets of malicious attacks on mobile devices. As noted by the founding researchers of the dataset; approximately 80\% of the world’s network traffic is generated by mobile devices~\cite{lashkari2018toward}. Consequently, attackers are targeting mobile devices at an increasing rate, and it is imperative that researchers have a representative dataset tailored specifically to this scenario in order to develop effective mitigating strategies. According to a 2017 report by Symantec, the ratio of benign to malicious software on mobile devices is 80/20~\cite{Kavitha_2017}. This represents a rather large proportion of malicious software and emphasises the need for datasets tailored specifically for this use case. The CICAndMal2017 dataset is one of the most comprehensive efforts toward this goal. It is comprised of 42 types of malicious software across 4 broad categories – adware, ransomware, scareware, and SMS malware along with benign records. Furthermore, the dataset is comprised of 80 traffic features to allow researchers to conduct their own feature selection as per their specific application. The testbed architecture of the dataset is described next.

\begin{figure}[htp]
    \centering
    \includegraphics[width=\columnwidth]{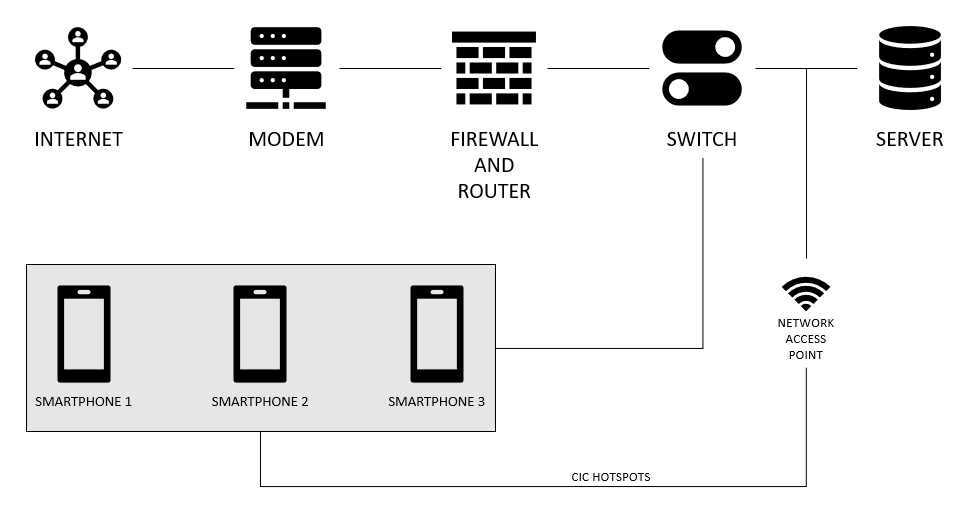}
    \caption{System architecture}
    \label{fig:cic_andmal}
\end{figure}

Three Android smartphones (connected to laptops, to automate certain tasks) were used for the data generation process. 429 malware and 5065 benign apps were installed across the smartphones and the network traffic generated by each was recorded. An emphasis was placed on recording traffic generated in three specific cases (based on prior knowledge of malware apps): immediately after installation, 15 minutes before the phones were rebooted and 15 minutes after the phones were rebooted. The system architecture is illustrated in Figure~\ref{fig:cic_andmal}.

\textbf{Related Literature:} The 2019 study by Noorbehbahani et al.~\cite{noorbehbahani2019analysis} was one of the first studies that employed this dataset. The researchers were concerned with the evaluation of various machine learning techniques for ransomware detection and cite two reasons for employing the CICAndMal2017 dataset for their analysis. First, the fact that this dataset was generated using real smartphones and not emulators serves to boost researchers’ confidence in its validity and reliability. The other contributing factor for this choice is that the dataset is large and comprehensive enough that even though the researchers were interested in investigating only 1 of the 4 categories of malware captured in the dataset, there was enough volume within that one category to still be useful for the intended purpose. Both these factors are indicative of the value of the CICAndMal2017 dataset and the kind of questions that it allows researchers to investigate. Based on their experiments with the CICAndMal2017 dataset, the researchers of~\cite{noorbehbahani2019analysis} concluded that amongst various machine learning techniques, random forest based models seem to be the most effective at identifying ransomware.

The 2021 study by Gohari et al.~\cite{gohari2021android} was concerned with evaluating the efficacy of deep learning techniques for identifying malware on Android smartphones. The researchers also cite CICAndMal2017 dataset’s generation using real smartphones as a major factor that informed their decision regarding the dataset to be employed, similar to the researchers discussed above. Furthermore, the comprehensive nature of the dataset – the wide variety of malware represented as well as the inclusion of 80 traffic features make it suitable for unsupervised deep learning techniques where the algorithm need not be fed labelled training data and can perform relevant feature selection on its own. Based on their experiments with the CICAndMal2017 dataset, the researchers in~\cite{gohari2021android} were able to create a CNN-LSTM deep learning model that achieved a malware category classification precision of 98.9\% and a malware family classification precision of 97.29\%. The recent publication of this study is an encouraging sign of the research opportunities that have been enabled by the CICAndMal2017 dataset and indicates that it will continue inform research endeavours for the foreseeable future.

\subsubsection{Bot-IoT}
\textbf{Overview:} The rise of IoT has created a new avenue that can be exploited by cyber attackers. Due to the unique nature of the IoT, it is imperative to develop modern datasets that are representative of the nature of traffic flow on such networks and can therefore be used to study and evaluate the potential of novel intrusion detection techniques aimed specifically at such networks. The Bot-IoT dataset is one effort in this direction. The dataset was published in 2018 by researchers at UNSW Canberra~\cite{koroniotis2019towards} and was created by simulating IoT device traffic through virtual machines. Specifically, several virtual machines were instantiated and used to simulate IoT sensors connected to a public IoT hub.  The types of sensors that were simulated are described below~\cite{koroniotis2019towards}:

\begin{enumerate}
    \item Weather station generating datapoints on air pressure, humidity, and temperature.
    \item Smart refrigerator that measures and regulates its temperature as required.
    \item “Motion-activated lights which turn on or off based on a pseudo-random generated signal.”
    \item “Remote-controlled garage door that opens and closes based on a probabilistic input.”
    \item Smart thermostat that measures the temperature of a house and regulates it as required by turning the air conditioning on or off.
\end{enumerate}

By simulating these devices, the researchers were able to create over 72 million records of IoT traffic data that is representative of the kind of traffic that can be observed on modern networks. Furthermore, the dataset contains several attacks that can be encountered on a modern IoT network such as: DDoS, DoS, OS and service scan, Keylogging and Data exfiltration. It is clear from our discussion that the Bot-IoT dataset is a comprehensive dataset that is representative of modern IoT networks and should serve as a much better benchmark to evaluate the performance of novel IDS techniques envisioned for IoT networks rather than older datasets that were not developed for the IoT and thus may not be representative of the nature of traffic and attacks that are involved in such networks.

\textbf{Related Literature:}
Since its publication in 2018, this dataset has been widely adopted for evaluation of IDS techniques developed specifically for IoT networks. Let us examine two recent studies that have used the Bot-IoT dataset in ways that serve as examples of its validity and wide applicability.
The first such study~\cite{ibitoye2019analyzing} evaluates the performance of two deep learning techniques – Feed Forward Neural Networks (FNN) and Self-Normalizing Neural Networks (SNN), by deriving adversarial samples from the Bot-IoT dataset using IBM’s Adversarial Robustness Toolbox. The dataset’s “realistic representation”~\cite{ibitoye2019analyzing} of IoT networks as well as its large size of heterogenous and malicious records afford it a level of malleability that was previously unattainable; even more so for newer network architectures such as that of IoT devices. By deriving adversarial samples, the researchers were able to demonstrate that even with the immensely popular FNNs there is a significant drop in performance when evaluated solely on adversarial samples. Specifically, the tested FNN achieved an accuracy of 95.1\% on the standard data but an average accuracy of only 24.33\% on adversarial samples~\cite{ibitoye2019analyzing}. Their proposed SNN was able to bridge the gap partially but also performed considerably worse on adversarial samples. This result highlights the need for more sophisticated IDS techniques to be developed and tested on an increasing number of adversarial samples – a finding that would have been impossible without the Bot-IoT dataset.

The next study that we will examine leverages the Bot-IoT dataset’s large number of available features to deploy an information gain based strategy for the feature selection of a hybrid IDS technique~\cite{khraisat2019novel}. The study uses information gain to effectively identify the most significant features and uses them to build a hybrid IDS that leverages both signature-based and anomaly-based techniques. The anomaly-based techniques were used to recognize zero-day attacks and signature-based methods were used to identify well established attacks. The results of the study are encouraging – the used methodology can attain significant improvement in detection accuracy and reveal the benefits of both effective feature selection and hybrid IDS techniques. The contributions of this study with respect to the importance of feature selection is powered by the reliability and variety of the Bot-IoT dataset and should serve as motivation to develop even more comprehensive, reliable, and large-scale datasets to empower cybersecurity research.

\subsubsection{Network TON\_IoT}

\textbf{Overview:} This one of the most recent datasets that has been created specifically for IoT networks. The dataset was published in 2021 and the motivation behind its development was the lack of credible datasets for applications specific to IoT networks. The concerned researchers cite the “heterogeneity, complexity and non-standardization of IoT networks”~\cite{moustafa2021new} as one of the main reasons for this discrepancy. To overcome this challenge, the researchers propose a testbed architecture that is comprised of three layers that comprise modern IoT networks – edge, fog, and cloud. The configuration and purpose of each of these layers is discussed below:

\begin{enumerate}
\item Edge: This layer consists of the end-user facing devices that generate raw data. These include devices like routers, laptops and IoT devices like weather sensors, smart lights, etc.
\item Fog: This layer differs from the edge layer in that it is not located at the point of data collection, but rather constitutes the local area network on which edge devices communicate. Thus, this layer is comprised of devices like servers and data management systems that are responsible for the smooth operation of user devices. This layer implements Software Defined Networks (SDN), Service Orchestration (SO) and Network Function Virtualization (NFV)~\cite{moustafa2021new}.
\item Cloud: This layer consists of cloud services such as the HIVE MQTT dashboard that enables the collection and analysis of telemetry data from sensor devices and cloud data analysis services such as Microsoft Azure and Amazon Web Services.
\end{enumerate}

Using this architecture, the researchers were able to develop a comprehensive dataset that represents the nature of benign and malicious traffic on IoT networks. The following attack types are represented in the dataset: Scanning attack, DoS, DDoS, Ransomware, Backdoor, Injection, Cross-site scripting, Password cracking and Man-In-The-Middle.

\textbf{Related Literature:} Since this is an extremely recent dataset, we were unable to find any research papers that have used this dataset for model evaluation. However, the researchers that have developed the dataset also tested the performance of several machine learning models on it in~\cite{moustafa2021new}. These results are not indicative of actual performance and have not been discussed extensively in this study because the researchers included attributes like the IP addresses and ports that can significantly boost detection accuracy. The researchers have also noted this caveat in~\cite{moustafa2021new} and have recommended that future endeavours implement models without using these features to ensure that the dataset represents the complexity and patterns of real-world data. This dataset has been included in our discussion because despite its lack of implementation at the time of writing, it represents one of the most comprehensive and versatile datasets developed specifically for applications within IoT and serves as an excellent example for future datasets in other branching fields to draw inspiration from.

\subsubsection{CIC DoS}
\textbf{Overview:} This dataset was published in 2017 by researchers at the Canadian Institute for Cybersecurity~\cite{jazi2017detecting}. The development of the dataset was motivated by the lack of suitable datasets for the development and evaluation of detection techniques for application layer DoS attacks. Denial of Service (DoS) attacks are one of the earliest kinds of intrusions and have taken several forms as the information technology landscape has evolved. Application layer DoS attacks are one such variation of the traditional network layer DoS attacks that by “Focusing on specific characteristics and vulnerabilities of application layer protocols, are capable of inflicting the same level of impact as traditional flooding DoS attacks at a much lower cost.”~\cite{jazi2017detecting}. The researchers in~\cite{jazi2017detecting} were concerned with investigating these application layer DoS attacks and decided to create a new dataset for evaluation of their detection techniques due to the lack of suitable datasets.

The researchers set up a victim server and generated attacks modelled after the most common types of application layer DoS attacks. Certain assumptions were made about the attacker that should be valid in most real-world scenarios as well:

\begin{enumerate}
\item “The attacker knows exactly when and how much traffic to send to maximize attack damage.”~\cite{jazi2017detecting}
\item The attack was stopped as soon as it was apparent that the attack had been successful –   indicated by the fact that the victim server had become unresponsive.
\end{enumerate}

Finally, the generated attacks were combined with the benign traffic records from the 2012 ISCX dataset to generate the final CIC DoS dataset. This is an excellent example of how new datasets can be generated in a timely and cost-effective fashion for specific use cases by leveraging parts of older datasets that are deemed to be relevant.

\textbf{Related Literature:} The publication of this dataset has prompted other researchers to investigate application layer DoS attacks and provided them with an effective tool to evaluate proposed detection techniques. Researchers in~\cite{perez2020flexible} were interested in investigating a specific kind of application layer DoS attack – Low Rate DDoS attacks (LR-DDoS). The evaluation of detection techniques specifically for LR-DDoS attacks would not have been possible without the CIC DoS dataset as it is the only one – to the best of our knowledge, that covers this class of attacks. Using this dataset, researchers were able to develop various machine learning based detection techniques and achieve a detection rate of 95\%. This is an exceptional achievement driven by the CIC DoS dataset because application layer DoS attacks including LR-DDoS attacks have been notorious in the cybersecurity community for being stealthy and especially hard to detect.

Another study published in 2019~\cite{ccalicsir2019intrusion} leveraged the CIC DoS dataset to determine the most suitable machine learning techniques for detecting application layer DoS attacks. As stated earlier, the creation of the CIC DoS dataset has encouraged research endeavours in the relatively understudied area of application layer DoS attacks and should serve as an example to the research community at large. It is an exhibition of the fact that new and valuable datasets in emerging fields can be created in a timely and cost-effective manner by leveraging older datasets instead of generating entirely new datasets from scratch. The researchers in~\cite{ccalicsir2019intrusion} experimented with several machine learning models and after evaluating the performance of each on the CIC DoS dataset, came to the conclusion that Light Gradient Boosting Machine (LGBM) methods are the most effective at detecting application layer DoS attacks.

\subsubsection{CICIDS2017}

\textbf{Overview:} This dataset was published in 2017 as a solution to the lack of modern datasets for developing and evaluating novel intrusion detection techniques. It covers seven major modern attack types and is available publicly to aid the research and development efforts in the area. The architecture of the testbed is described below~\cite{sharafaldin2018toward}:

The dataset has been created by implementing two distinct networks: a victim network and an attacker network. The victim network has been designed such that it incorporates elements of typical modern networks, specifically, the network consists of three servers, a router, a firewall, two switches and ten computers interconnected through a domain controller and active directory. Furthermore, the computers use all three of the most popular operating systems – Windows, Linux, and Mac. The attacker network is configured such that it comprises of one router, one switch and four computers running the Kali Linux and Windows 8.1 operating systems.

One of the major attributes of this dataset that make it extremely useful in modern scenarios is its attention to generating real world like benign traffic. The researchers developed a novel profiling system~\cite{sharafaldin2018towards} that uses machine learning to analyse how human behaviour and interaction is abstracted on connected networks to generate benign traffic that simulates organic traffic as closely as possible. The created dataset contains traffic recorded over a five-day period, instances of seven modern attacks (Brute force, Heartbleed, Botnet, DoS, DDoS, Web, and Infiltration) and is comprised of 80 network flow features.

\textbf{Related Literature:} Let us now examine some recent research papers that have used the CICIDS2017 dataset to investigate its suitability and reliability to evaluate modern IDS techniques.
The first paper we will discuss is~\cite{roopak2019deep}. This work is concerned with developing a deep learning-based IDS technique for IoT networks and uses the CICIDS2017 dataset to evaluate the model’s performance. It is an encouraging fact that even though the CICIDS2017 dataset was not created specifically for IoT networks, it can be used in this context – providing evidence for the wide range of attacks represented and possible applications. The research paper was published in 2019 and the researchers cite the dataset’s resemblance to real-world networks as the reason why CICIDS2017 was used for evaluation. The researchers implemented several deep learning-based models, focussing primarily on DDoS attacks and obtained the highest precision of 98.44\% through an LSTM model. Their worst performing model was an MLP that achieved precision of 88.47\%. These performance metrics are relatively high and represent exceptional intrusion detection performance. These results are encouraging because since the CICIDS2017 dataset is a reliable representation of real-world network traffic, we can conclude that deep learning techniques provide an effective tool for intrusion detection and should be developed further.

Another study that uses the CICIDS2017 dataset for model evaluation is~\cite{ahmim2019novel}. This study is unique in that it not only uses CICIDS2017 to evaluate the performance of the proposed technique but also evaluates the performance of historical models, going as far back as 2009, on the same dataset to enable fair comparisons. The fact that the dataset can be used in this context to evaluate the performance of models that were created almost a decade before the dataset is a testament to the wide variety and all-encompassing nature of network traffic in this dataset. It is an important feature that should inform the creation of future datasets. A modern dataset should not only capture the intricacies of modern network traffic prevalent at the time of creation but should also incorporate enough historical patterns such that the relevance of older models can be evaluated and the efficacy of modern IDS techniques against older attack types can be confirmed.

\subsubsection{UCSD Network Telescope}

\textbf{Overview:} The last dataset that we will discuss is The UCSD Network Telescope. Even though this dataset was publsihed earlier than 2017, it merits inclusion in our analysis because it is a dynamic dataset that is constantly evolving and is unique in its nature and sheer scale. It “consists of a globally routed, but lightly utilized /8 network prefix”~\cite{UCSD_2013} and is one of the largest sources of private traffic in the world – capturing 1/256th of the entire IPv4 address space. The dataset is comprised of traffic traces and is made available in near real-time in the form of one hour long compressed pcap files. The telescope has been in operation since 2008 and collects over 3TB of data every day. This is one of the largest projects of its kind and has been leveraged by researchers to generate insights that would not have been possible without such data. Two recent examples of such research endeavours are discussed next to emphasize the value created by this dataset and the importance of establishing more such resources to bolster research efforts around the world.

\textbf{Related Literature:} The emergence and widespread adoption of IoT devices has inadvertently enabled the execution of large-scale cyber-attacks by exploiting a large number of small IoT devices. The 2020 study by Torabi et al.~\cite{torabi2020scalable} proposed an innovative solution to this problem. By leveraging Big Data technologies and the UCSD Network Telescope data, the researchers were able to develop a tool that can monitor network traffic and identify compromised IoT devices that may be used for executing cyber-attacks. The tool was able to process over 4TB of passive network data captured by the UCSD Network Telescope to “to identify 27,849 compromised IoT devices that generated more than 300 million unsolicited packets”~\cite{torabi2020scalable}.

The 2020 study by Lutscher et al.~\cite{lutscher2020home} is an example of how cybersecurity datasets can be used to inform interdisciplinary research and generate valuable insights in other disciplines as well. The researchers in this study were interested in the social and political impact of DoS attacks. Specifically, the researchers posited that countries with nondemocratic forms of government would be more prone to DoS attacks around the time of politically important events such as elections because such attacks are used by used by governments to “censor regime-threatening information” and by activists to “publicly undermine the government’s authority”~\cite{lutscher2020home}. Historical passive network data from the UCSD Network Telescope was used to identify various DoS attacks across the world and the results indicated that indeed, the number of DoS attacks in authoritarian countries spiked around the time of elections. However, the attacks were not directed internally within the country but rather externally at institutions and organisations around the world that would serve as news resources regarding the concerned country.


\section{Taxonomy of recent datasets}
\begin{figure}[htp]
    \centering
    \includegraphics[width=\columnwidth]{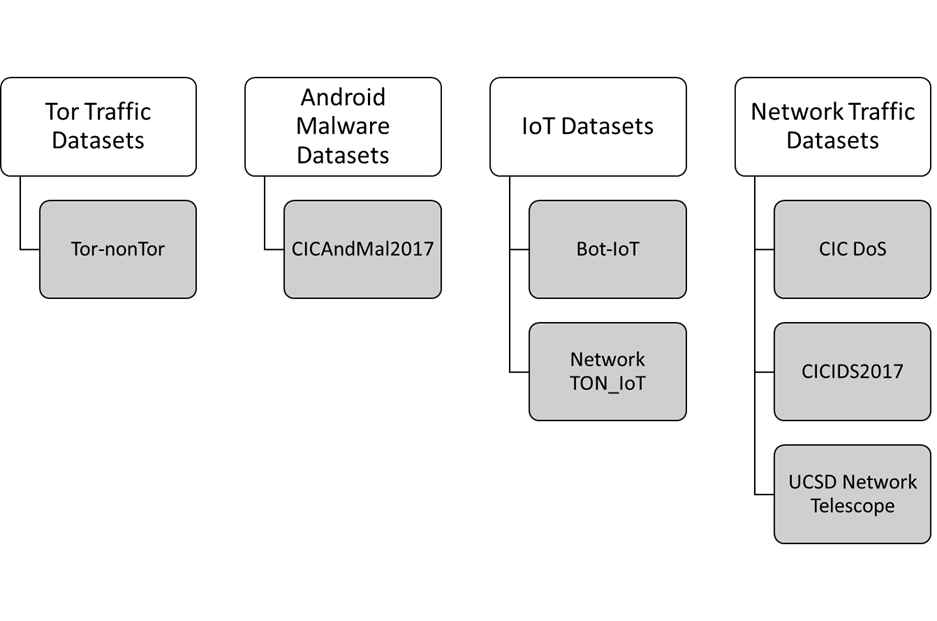}
    \caption{Taxonomy of recently developed datasets}
    \label{fig:my_label}
\end{figure}

Given the wide range of environments and applications of IDS datasets, it is a worthwhile endeavour to develop a taxonomy that can help categorize and organize datasets such that high-level comparisons can be made and an appropriate set of datasets as per the research objective can be shortlisted. In the current analysis, we have created a taxonomy (Figure~\ref{fig:my_label}) based on the IT environment that the dataset has been generated from and classified each of the 7 discussed datasets into 4 broad classes:

\begin{enumerate}
    \item Tor Traffic Datasets: These datasets have been generated from traffic on the Tor network. The primary application of such datasets is for privacy preservation research in online communications and decoding patterns in the traffic to identify illegal activity.
    \item Android Malware Datasets: These datasets pertain to traffic generated by smartphone applications. As per Google, over 50\% of internet traffic is generated by smartphone devices~\cite{google}. Therefore, it is imperative for the IDS research community to study this environment and develop effective solutions to ensure that the security and privacy of smartphone users is protected.
    \item IoT datasets: Another novel type of communications device that is becoming ubiquitous across modern networks is the internet of things. It is estimated that there will be 43 billion IoT devices in the world by 2023~\cite{dahlqvist_patel_rajko_shulman_2020}. This far exceeds the number of any other communications device on the planet and provides an expansive new frontier for cyber criminals to exploit. Therefore, the development of IoT traffic datasets to aid IDS research in the area is of utmost importance and should arguably be the number one priority of the cyber security research community.
    \item Network Traffic Datasets: This class refers to datasets that have been generated from traditional network traffic. Even older datasets such as the NSL-KDD dataset would be classified under this category, so there are plenty of datasets to choose from in this category. However, given the dynamic nature of the cyber threats landscape, the development of new datasets that accurately encapsulate the nature of attacks on modern networks is still of immense value to the research community.
\end{enumerate}

Furthermore, the discussed datasets are summarized below in Table~\ref{tab:my-table} to indicate their categorization in the taxonomy, their attributes and their ideal use-cases. It is apparent from the discussion above that the development of new datasets that represent the nature of traffic across newer technologies like smartphones and IoT, as well as the ever-evolving nature of threats on modern networks is imperative to aid research endeavours that can help ensure the highest standard of security and privacy for users across all forms of networked communications.

\FloatBarrier
\begin{table}[]
\footnotesize
\centering
\begin{tabular}{@{}|p{2cm}|p{3cm}|p{3cm}|p{4cm}|@{}}
\toprule
\textbf{Category} &
  \textbf{Dataset} &
  \textbf{Number of attributes} &
  \textbf{Ideal use case} \\ \midrule
Tor traffic &
  Tor-nonTor &
  8 types of traffic from more   than 18 applications &
  Privacy preservation techniques   for online communication \\ \midrule
Android malware &
  CICAndMal2017 &
  4 broad malware categories and   more than 80 network flow attributes &
  Analysis of malware on modern   smartphones \\ \midrule
\multirow{2}{*}{IoT} &
  Bot-IoT &
  72mn records across 6 attack   types &
  Developing data intensive models   for IoT intrusion detection \\ \cmidrule(l){2-4} 
 &
  Network TON\_IoT &
  9 attack types &
  Developing versatile models for   IoT networks than can identify wide gamut of attacks \\ \midrule
\multirow{3}{*}{Network traffic} &
  CIC DoS &
  4 types of attacks &
  Development of detection   techniques for application layer DoS attacks \\ \cmidrule(l){2-4} 
 &
  CICIDS2017 &
  80 network flow attributes &
  Developing versatile techniques   for detection of wide range of intrusions \\ \cmidrule(l){2-4} 
 &
  UCSD Network Telescope &
  1/256th of the entire IPv4   address space &
  Developing data intensive models   to uncover patterns in global network communication \\ \bottomrule
\end{tabular}%
\caption{Summary of discussed datasets}
\label{tab:my-table}
\end{table}
\FloatBarrier

\FloatBarrier
\FloatBarrier

\section{Conclusion and directions for future research}
In this survey we conducted an in-depth review of 7 datasets that have been published recently for intrusion detection research by reputable institutions in the research community. Through our discussions we highlighted the rapid rate of development of the information technology industry and the evolution of branching industries such as IoT and smartphone devices to such an extent where traditional IDS datasets are no longer applicable for these specific use-cases. By exploring the solutions developed on the back of the considered modern datasets, we have shed light on the need for the development of even more datasets that are representative of the modern IT environment to ensure that intrusion detection research endeavours are not limited by the lack of appropriate evaluation datasets. Specifically, future efforts should aim to develop even more datasets for underrepresented branching technologies such as IoT, smartphones, wearable sensor devices, etc. for traffic based IDS. Furthermore, there is a pressing need for more datasets to be published to enable research endeavours concerned with privacy-preservation in an increasingly online world.

\clearpage
\bibliographystyle{IEEEtran}
\bibliography{IEEEabrv, bibliography.bib}
\end{document}